\documentclass{article}

\usepackage{arxiv}

\usepackage[utf8]{inputenc} 
\usepackage[T1]{fontenc}    
\usepackage{hyperref}       
\usepackage{url}            
\usepackage{booktabs}       
\usepackage{amsfonts}       
\usepackage{nicefrac}       
\usepackage{microtype}      
\usepackage{lipsum}
\usepackage{graphicx}
\graphicspath{ {./images/} }
\usepackage{amsmath}
\usepackage{bm}
\usepackage{amssymb,amsthm,amsmath}
\usepackage{xcolor,paralist,hyperref,titlesec,fancyhdr,etoolbox}
\usepackage{algorithm}
\usepackage{algpseudocode}
\usepackage{makecell}
\newtheorem{theorem}{Theorem}[section]

\newtheorem{remark}[theorem]{Remark}
\newtheorem{assumption}{Assumption}

\title{Koopman-Based Model Predictive Control for Simultaneous State of Charge and Temperature Balancing of Lithium-Ion Cells}

\author{
 Jihoon Moon \\
  Department of Mechanical Engineering\\
  The Pennsylvania State University\\
  University Park, PA 16802 \\
  \texttt{jihoonmoon@psu.edu} \\
}

\begin{document}
\maketitle
\begin{abstract}
Cell-to-cell electrical and thermal variations produce nonuniform state of charge (SoC) and temperature distributions in lithium-ion battery packs. This paper proposes a Koopman-operator-based framework for simultaneous SoC and temperature balancing of series-connected cells. Cell-specific Koopman predictors identified using extended dynamic mode decomposition are assembled into neighboring-cell error dynamics that retain heterogeneity and thermal-disturbance effects. A Tikhonov-regularized feedforward controller provides numerically robust compensation without excessive current amplification, while constrained Koopman-based model predictive control (KMPC) regulates the residual errors through a convex quadratic program. Three-cell simulations validate the controller under two consecutive Urban Dynamometer Driving Schedule cycles. KMPC achieves balancing performance comparable to nonlinear model predictive control (NMPC). Relative to the uncontrolled case, KMPC reduces the RMSEs of the neighboring-cell SoC and core temperature differences by $66$--$81\%$ and $57$--$76\%$, respectively, while reducing the computation time per optimization from $0.0259~\mathrm{s}$ for NMPC to $0.0028~\mathrm{s}$ for KMPC.
\end{abstract}


\section{Introduction}
Lithium-ion batteries are widely used in electric vehicles and energy-storage systems because of their high energy density and efficiency. Battery packs for these applications typically comprise numerous cells whose characteristics inevitably differ because of manufacturing variability \cite{rumpf2018influence}. Even among cells from the same production batch, variations in key parameters may exceed 1$\%$ \cite{chen2022toward}. Such cell-to-cell variability produces nonuniform state-of-charge (SoC) and temperature distributions within a battery pack, thereby compromising operational safety and performance while accelerating degradation \cite{sorouri2025online,bhaskar2024post}. This variability may eventually lead to thermal runaway and therefore, require accurate safety assessment before catastrophic event occurs \cite{moon2025safe,moon2024short,moon2026detecting,moon2024state}. 

Electrical and thermal nonuniformity are interconnected. Differences in resistance, capacity, and thermal characteristics produce unequal current and heat generation \cite{dubarry2009single,paul2013analysis,ahuja2026lithium}. The resulting temperature gradients alter reaction kinetics, open-circuit voltage, resistance, and aging rate, which in turn amplify electrical mismatch. Cell-level temperature difference of 3$^{\circ}$C-5$^{\circ}$C can induce substantial current nonuniformity, exacerbate existing imbalances, and shorten battery cycle life \cite{rahn2013battery,docimo2018analysis}. Moreover, because cells subjected to different electrical and thermal conditions degrade at different rates, these imbalances tend to become more pronounced as the battery pack ages \cite{paul2013analysis}. Therefore, since the electrical and thermal nonuniformity are tightly coupled and may conflict under a shared current input, the simultaneous thermal and SoC balancing is needed \cite{altaf2014simultaneous}.

To mitigate cell heterogeneity, numerous cell balancing methods are introduced. A multilevel-converter-based modular battery architecture employed model predictive control (MPC) to allocate the load among cells while tracking terminal voltage and balancing both SoC and temperature \cite{altaf2016load}. The tradeoff between charge and temperature imblanace is formalized and demonstated that a balancing action that improves one state can worsen the other under certain operating conditions \cite{docimo2018analysis}. Distributed multi-objective consensus control has been used to coordinate SoC, temperature, power capability, and loss-related objectives with improved scalability and reduced dependence on a centralized controller \cite{barreras2021consensus}. These studies establish active current allocation as a viable means of electro-thermal balancing, but also show that performance depends critically on the predictive model used by the controller.

Recent work has expanded the treatment of heterogeneity and nonlinear control. Linear time-varying heterogeneity models support joint estimation and balancing of multiple cell-state differences \cite{docimo2022estimation}, while parameter-aware analysis shows that capacity mismatch can continually
regenerate SoC error under nonzero pack current unless the balancing law compensates for cell-specific dynamics \cite{abadie2022framework}. Aging-aware nonlinear optimization has enforced SoC, current, and temperature limits for cells with different states of health \cite{duy2025optimal}, and adaptive
electro-thermal weighting has been investigated using particle-swarm optimization \cite{cao2025coordinated}. Learning-based policies provide another route to multi-objective control: reinforcement learning has been experimentally demonstrated for concurrent SoC and temperature balancing \cite{li2025reinforcement}, and online brain-emotional-learning control has shown
robustness to operating variability and measurement uncertainty \cite{sorouri2025online}.

Despite this progress, a persistent gap remains between nonlinear electro-thermal fidelity and a model form that permits transparent, computationally predictable control synthesis. Simplified or locally linearized models may lose accuracy across broad SoC and temperature ranges, whereas nonlinear MPC can require a nonconvex optimization online. Learning-based policies can reduce online computation after training, but their behavior depends on the training distribution, reward design, and hyperparameter selection. Therefore, a control-oriented representation is needed that captures nonlinear coupling and heterogeneity while preserving the linear structure required for efficient optimization.

This paper develops a Koopman-operator-based framework for simultaneous SoC and temperature balancing of series-connected lithium-ion cells equipped with bidirectional active-balancing converters. Cell-specific Koopman predictors are identified from heterogeneous nonlinear electro-thermal data using extended dynamic mode decomposition and assembled into neighboring-cell error dynamics that capture parameter mismatch, nominal-current, and thermal-disturbance effects. A Tikhonov-regularized feedforward controller provides numerically robust compensation without excessive balancing currents, while constrained Koopman-based MPC drives the residual errors toward consensus through a convex quadratic program. Three-cell simulations validate the predictors under an unseen current profile and demonstrate balancing performance comparable to nonlinear MPC over repeated Urban Dynamometer Driving Schedule cycles, with substantially lower online computation time.

\section{Preliminaries}
\subsection{Active-Balancing Hardware for Series-Connected Cells}
\begin{figure}
    \centering
    \includegraphics[width=100mm]{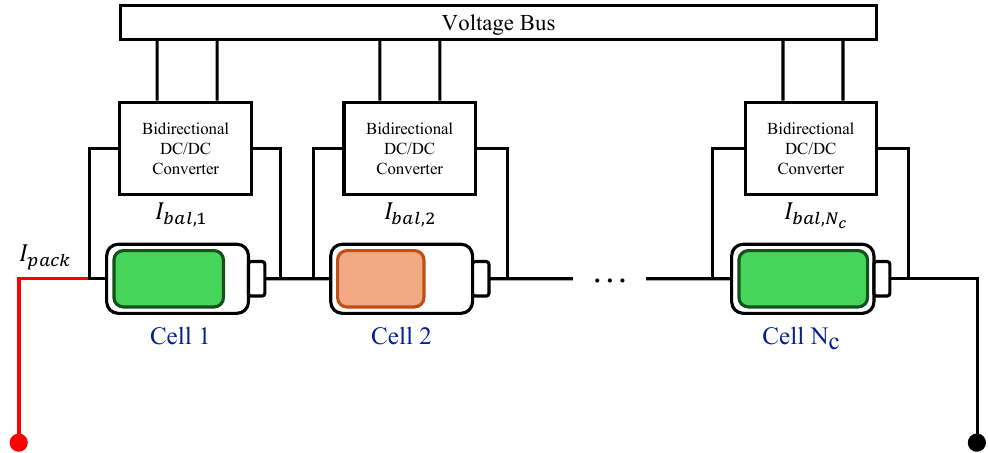}
    \caption{Battery pack configuration.}
    \label{figure_1}
\end{figure}

The series-connected cells are interfaced with bidirectional active-balancing DC/DC converters (Fig. \ref{figure_1}), enabling independent regulation of the balancing current for each cell. Compared with passive balancing methods, this architecture can accelerate the balancing process by allowing larger balancing currents to flow between cells \cite{turksoy2020comprehensive}.

\subsection{Koopman Operator Theory \cite{mezic2005spectral}}
Consider an autonomous discrete-time dynamical system
\begin{equation}
	\bm{x}(k+1)=f(\bm{x}(k)),
\end{equation}
where $\bm{x}(k)\in\mathcal{M}\subseteq\mathbb{R}^{N_x}$ is the system state at sampling instant $k$, and $f:\mathcal{M}\rightarrow\mathcal{M}$ denotes the nonlinear state-transition map. Rather than propagating the state directly, Koopman operator theory describes the evolution of observable functions. Let $\mathcal{F}$ be a function space whose scalar elements $\psi:\mathcal{M}\rightarrow\mathbb{R}$ remain in $\mathcal{F}$ after composition with $f$. The Koopman operator $\mathcal{K}:\mathcal{F}\rightarrow\mathcal{F}$ is then defined by
\begin{equation}
	(\mathcal{K}\psi)(\bm{x}(k))=\psi\!\left(f(\bm{x}(k))\right). \label{Koopman}
\end{equation}
Using the state-update equation, the action of $\mathcal{K}$ can equivalently be written as
\begin{equation}
	(\mathcal{K}\psi)(\bm{x}(k))=\psi(\bm{x}(k+1)).
\end{equation}
Although the state-transition map $f$ may be nonlinear, $\mathcal{K}$ evolves observables linearly. An exact Koopman representation is generally infinite-dimensional; therefore, data-driven implementations such as extended dynamic mode decomposition (EDMD) typically construct a finite-dimensional approximation using a selected dictionary of observable functions \cite{ren2025koopman}.

\subsection{Koopman Operator with Control}
For a control-affine nonlinear system with a control input $\bm{u}(k)\in\mathbb{R}^{N_u}$,
\begin{equation}
	\bm{x}(k+1)=f\!\left(\bm{x}(k)\right)+g\!\left(\bm{x}(k)\right)\bm{u}(k),
\end{equation}
EDMD approximates the controlled Koopman operator in a finite-dimensional observable space \cite{korda2018linear}. Given a lifting map $\bm{\Psi}:\mathcal{M}\rightarrow\mathbb{R}^{N_o}$, the lifted state is defined as
\begin{equation}
	\bm{z}(k)=\bm{\Psi}\!\left(\bm{x}(k)\right),
\end{equation}
where $\bm{z}(k)=[\psi_1(\bm{x}(k)), \psi_2(\bm{x}(k)),\ldots,\psi_{N_o}(\bm{x}(k))]^\top \in \mathbb{R}^{N_o}$ and $N_o$ denotes the number of observables. The finite-dimensional Koopman system matrices are identified from measured state--input data by solving
\begin{equation}
	\min_{A,B}\sum_{k=1}^{N_s}
	\left\|\bm{\Psi}\!\left(\bm{x}(k+1)\right)
	-A\bm{\Psi}\!\left(\bm{x}(k)\right)-B\bm{u}(k)\right\|_2^2, \label{edmd}
\end{equation}
where $N_s$ is the number of state-input transition samples used for model identification. This approximation yields the linear lifted predictor
\begin{equation}
	\bm{z}(k+1)=A\bm{z}(k)+B\bm{u}(k),
	\qquad \hat{\bm{x}}(k)=C\bm{z}(k),
\end{equation}
where $A\in\mathbb{R}^{N_o\times N_o}$, $B\in\mathbb{R}^{N_o\times N_u}$, and $C\in\mathbb{R}^{N_x\times N_o}$. The matrix $C$ maps the lifted state back to the original state coordinates and may be selected directly when those states are included in $\bm{\Psi}$.

\section{Electro-Thermal Model}
Cell-to-cell variability can arise from differences in the electrical and thermal parameters of individual cells \cite{rumpf2017experimental,fill2022experimental}. Accordingly, a heterogeneous electro-thermal model with cell-specific parameters is adopted to represent this nonuniform behavior. To distinguish the balancing-control problem from parameter-estimation uncertainty, the following assumption is introduced.
\begin{assumption}\label{assump1}
The cell-specific electrical and thermal parameters are known and remain constant over the operating horizon considered in this study.
\end{assumption}
\begin{remark}
	Assumption~\ref{assump1} isolates the balancing-control design from errors associated with parameter estimation. In practical battery-management applications, the cell-specific parameters are provided through offline characterization or online identification. Robustness to parameter uncertainty is outside the scope of this study.
\end{remark}

The heterogeneous electro-thermal model is employed to describe the dynamics of each lithium-ion battery (LIB) cell, where the subscript $i$ identifies an individual cell. The electrical subsystem is represented by a first-order Thevenin equivalent circuit model (ECM), whose states are the state of charge $SoC$ and polarization voltage $V_c$, with terminal voltage $V$ as the output. The corresponding discrete-time model is given by
\begin{equation} 
    \begin{aligned} 
        SoC_i(k+1) &= SoC_i(k) -\dfrac{I_i(k)\Delta t}{36Q_{c,i}}, \\ 
        V_{c,i}(k+1) &= \left(1-\dfrac{\Delta t}{R_{1,i} C_{1,i}}\right)V_{c,i}(k) + \dfrac{I_i(k) \Delta t}{C_{1,i}}, \\
		V_{i}(k) &=  OCV(SoC_i(k))-V_{c,i}(k)-I_i(k)R_{0,i}.
\end{aligned}
\label{electrothermalmodel}
\end{equation}
The $SoC$ is updated using Coulomb counting, where $Q_c$ denotes the cell capacity. The $R_1$--$C_1$ branch characterizes polarization effects associated with slow diffusion processes. The terminal voltage is determined by the nonlinear open-circuit-voltage relationship $OCV(SoC)$, the polarization voltage $V_c$, and the ohmic voltage drop across the internal resistance $R_0$. Here, $\Delta t$ denotes the sampling interval, and the input current $I$ is defined as positive during discharge.

A two-state lumped-parameter thermal model is coupled with the ECM to describe the temperature dynamics of each cell. The thermal states are the core temperature $T_c$ and surface temperature $T_s$. The corresponding discrete-time model is expressed as
\begin{equation}
	\begin{aligned}
		T_{c,i}(k+1) ={}&
	        \frac{Q_{gen,i}(k)\Delta t}{C_{d,i}}+ \left(1-\frac{\Delta t}{C_{d,i}R_{d,i}}\right)T_{c,i}(k) + \frac{T_{s,i}(k)\Delta t}{C_{d,i}R_{d,i}}, \\
		T_{s,i}(k+1) ={}& T_{s,i}(k) + \Delta t\left(\frac{T_{f,i}(k)-T_{s,i}(k)}{C_{v,i}R_{v,i}} -\frac{T_{s,i}(k)-T_{c,i}(k)}{C_{v,i}R_{d,i}}\right). \\
	\end{aligned}\label{thermalmodel}
\end{equation}
In this formulation, the internal heat-generation rate is obtained as $Q_{gen,i}(k)=I_i(k)\left(OCV(SoC_i(k))-V_i(k)\right)$, and $T_{f}$ denotes the ambient temperature. The conductive thermal resistance $R_d$ characterizes heat transfer between the cell core and surface, whereas $R_v$ represents the effective convective resistance between the cell surface and its surroundings. The thermal capacitances $C_d$ and $C_v$ govern the transient responses of the core and surface temperatures, respectively \cite{lin2014lumped}.


\section{Koopman operator for Electro-Thermal Model}
The nonlinear $OCV$ relationship enters the heat-generation term and introduces strong coupling between the electrical and thermal dynamics. Although the electro-thermal model can be addressed using nonlinear control methods, such approaches may increase online computational requirements and complicate closed-loop stability analysis. Therefore, a finite-dimensional Koopman representation is constructed for each cell to obtain a tractable linear predictor suitable for controller synthesis.

The heterogeneous electro-thermal dynamics are written as
\begin{equation}
	\bm{x}_i(k+1) = f_i(\bm{x}_i(k),\bm{u}_i(k),\bm{d}_i(k)), \label{fx}
\end{equation}
where $\bm{x}_i=[SoC_i,V_{c,i},T_{c,i},T_{s,i}]^\top\in\mathbb{R}^{N_x}$ with $N_x=4$, $\bm{u}_i=I_i\in\mathbb{R}^{N_u}$, and $\bm{d}_i=T_{f,i}\in\mathbb{R}^{N_d}$ with $N_u=N_d=1$. The cell-specific map $f_i:\mathbb{R}^{N_x}\times\mathbb{R}^{N_u}\times\mathbb{R}^{N_d}\rightarrow\mathbb{R}^{N_x}$ contains the corresponding electrical and thermal parameters. A spatially uniform ambient condition is recovered by setting $\bm{d}_i=\bm{d}$ for all cells. Including the disturbance in the EDMD approximation gives
\begin{equation}
	\min_{A_i,B_i,E_i}\sum_{k=1}^{N_s}
	\left\|\bm{\Psi}\!\left(\bm{x}_i(k+1)\right)
	-A_i\bm{\Psi}\!\left(\bm{x}_i(k)\right)-B_i\bm{u}_i(k)-E_i\bm{d}_i(k)\right\|_2^2.
\end{equation}
Then, the linear Koopman operator is presented as
\begin{equation}
\begin{aligned}
	\bm{z}_i(k+1)&=A_i\bm{z}_i(k)+B_i\bm{u}_i(k)+E_i\bm{d}_i(k), \\
	\hat{\bm{x}}_i(k)&=C_i\bm{z}_i(k), 
\end{aligned} \label{K_lin}
\end{equation}
For the $N_s$ measured transitions of cell $i$, define the snapshot matrices as
\begin{equation}
\begin{aligned}
	X_i &= \left[\bm{x}_i(1),\ldots,\bm{x}_i(N_s-1)\right], \\
	Y_i &= \left[\bm{x}_i(2),\ldots,\bm{x}_i(N_s)\right], \\
	U_i &= \left[\bm{u}_i(1),\ldots,\bm{u}_i(N_s-1)\right], \\
	D_i &= \left[\bm{d}_i(1),\ldots,\bm{d}_i(N_s-1)\right].
\end{aligned}\label{snapshot}
\end{equation}
The corresponding lifted-state snapshot matrices are
\begin{equation}
\begin{aligned}
	X_{\mathrm{lift},i} &= \left[\bm{\Psi}(\bm{x}_i(1)),\ldots,
	\bm{\Psi}(\bm{x}_i(N_s-1))\right], \\
	Y_{\mathrm{lift},i} &= \left[\bm{\Psi}(\bm{x}_i(2)),\ldots,
	\bm{\Psi}(\bm{x}_i(N_s))\right].
\end{aligned}
\end{equation}
Thus, $X_i,Y_i\in\mathbb{R}^{N_x\times N_s-1}$, $U_i\in\mathbb{R}^{N_u\times N_s-1}$, $D_i\in\mathbb{R}^{N_d\times N_s-1}$, and $X_{\mathrm{lift},i},Y_{\mathrm{lift},i}\in\mathbb{R}^{N_o\times N_s-1}$. The least-squares estimates are
\begin{equation}
\left[A_i\;B_i\;E_i\right]
=Y_{\mathrm{lift},i}
\begin{bmatrix}X_{\mathrm{lift},i}\\U_i\\D_i\end{bmatrix}^{\dagger}, \qquad C_i=X_iX_{\mathrm{lift},i}^{\dagger},
\end{equation}
where $\dagger$ denotes the Moore--Penrose pseudoinverse.

\section{Error Dynamics}
To formulate the balancing problem, the heterogeneous-system error dynamics in \cite{zhang2025distributed} are adapted to the cell-specific Koopman models in \eqref{K_lin}. The control objective is to achieve consensus among neighboring cells,
\begin{equation}\label{state_consensus}
	\lim_{k\to\infty}
	\left\|\bm{x}_i(k)-\bm{x}_{i+1}(k)\right\|=0,
	\qquad i=1,\ldots,N_c-1,
\end{equation}
where $N_c$ is the number of cells. For clarity, consider the path spanning tree with edges $(i,i+1)$; the formulation extends to a general spanning tree through its incidence matrix. Define the lifted edge error as
\begin{equation}
	\bm{\delta}_i(k)=\bm{z}_i(k)-\bm{z}_{i+1}(k),
	\qquad i=1,\ldots,N_c-1.
\end{equation}
Using \eqref{K_lin}, the dynamics associated with each edge are
\begin{equation}
\begin{aligned}
	\bm{\delta}_i(k+1)&={}A_i\bm{\delta}_i(k)+B_i\bm{u}_i(k)-B_{i+1}\bm{u}_{i+1}(k) \\
	&+E_i\bm{d}_i(k)-E_{i+1}\bm{d}_{i+1}(k) +\left(A_i-A_{i+1}\right)\bm{z}_{i+1}(k).
\end{aligned}
\end{equation}
The global lifted error dynamics for all cells can then be written as
\begin{equation}\label{global_error_dynamics}
	\Delta(k+1)=\mathcal{A}_{\Delta}\Delta(k)
	+\mathcal{B}_{\Delta}\mathcal{U}(k)
	+\overline{\mathcal{A}}_{\Delta}\mathcal{Z}(k)
	+\mathcal{E}_{\Delta}\mathcal{D}(k),
\end{equation}
where
\begin{equation}
	\mathcal{A}_{\Delta}=\operatorname{diag}
	\left(A_1,A_2,\ldots,A_{N_c-1}\right),
\end{equation}
\begin{equation}
\mathcal{B}_{\Delta}=
\begingroup\setlength{\arraycolsep}{2pt}
\begin{bmatrix}
B_1 & -B_2 & 0 & \cdots & 0 \\
0 & B_2 & -B_3 & \cdots & \vdots \\
\vdots & \vdots & \vdots & \ddots & 0 \\
0 & \cdots & 0 & B_{N_c-1} & -B_{N_c}
\end{bmatrix}
\endgroup,
\end{equation}
\begin{equation}
\overline{\mathcal{A}}_{\Delta}=
\begingroup\setlength{\arraycolsep}{2pt}
\begin{bmatrix}
0 & A_1-A_2 & 0 & \cdots & 0 \\
0 & 0 & A_2-A_3 & \cdots & \vdots \\
\vdots & \vdots & \vdots & \ddots & 0 \\
0 & 0 & \cdots & 0 & A_{N_c-1}-A_{N_c}
\end{bmatrix}
\endgroup,
\end{equation}
and
\begin{equation}
\mathcal{E}_{\Delta}=
\begingroup\setlength{\arraycolsep}{2pt}
\begin{bmatrix}
E_1 & -E_2 & 0 & \cdots & 0 \\
0 & E_2 & -E_3 & \cdots & \vdots \\
\vdots & \vdots & \vdots & \ddots & 0 \\
0 & \cdots & 0 & E_{N_c-1} & -E_{N_c}
\end{bmatrix}
\endgroup.
\end{equation}
The stacked vectors in \eqref{global_error_dynamics} are
\begin{equation}
\begin{aligned}
	\Delta(k)&=
	\begin{bmatrix}\bm{\delta}_1^{\top}(k)&\cdots&
	\bm{\delta}_{N_c-1}^{\top}(k)\end{bmatrix}^{\top}, \\
	\mathcal{Z}(k)&=
	\begin{bmatrix}\bm{z}_1^{\top}(k)&\cdots&\bm{z}_{N_c}^{\top}(k)\end{bmatrix}^{\top}, \\
	\mathcal{U}(k)&=
	\begin{bmatrix}\bm{u}_1^{\top}(k)&\cdots&\bm{u}_{N_c}^{\top}(k)\end{bmatrix}^{\top}, \\
	\mathcal{D}(k)&=
	\begin{bmatrix}\bm{d}_1^{\top}(k)&\cdots&\bm{d}_{N_c}^{\top}(k)\end{bmatrix}^{\top}.
\end{aligned}
\end{equation}
Consequently, $\Delta\in\mathbb{R}^{(N_c-1)N_o}$, $\mathcal{Z}\in\mathbb{R}^{N_c N_o}$, $\mathcal{U}\in\mathbb{R}^{N_c N_u}$, and $\mathcal{D}\in\mathbb{R}^{N_c N_d}$. The corresponding matrices satisfy $\mathcal{A}_{\Delta}\in\mathbb{R}^{(N_c-1)N_o\times(N_c-1)N_o}$, $\mathcal{B}_{\Delta}\in\mathbb{R}^{(N_c-1)N_o\times N_c N_u}$, $\overline{\mathcal{A}}_{\Delta}\in\mathbb{R}^{(N_c-1)N_o\times N_c N_o}$, and $\mathcal{E}_{\Delta}\in\mathbb{R}^{(N_c-1)N_o\times N_c N_d}$.
When the lifting dictionary contains the original electro-thermal states, the reconstruction matrices can be chosen as a common state-selection matrix, $C_i=C$. The reconstructed physical-state error then satisfies $\hat{\bm{x}}_i(k)-\hat{\bm{x}}_{i+1}(k)=C\bm{\delta}_i(k)$.


 

\section{Controller Design}
\subsection{Composite Controller Design}
A composite input is introduced to compensate for the heterogeneous-dynamics $\overline{\mathcal{A}}_{\Delta}\mathcal{Z}(k)$ and thermal-disturbance terms $\mathcal{E}_{\Delta}\mathcal{D}(k)$ in \eqref{global_error_dynamics} while retaining a feedback component for cell balancing. The stacked input is decomposed as
\begin{equation}
	\mathcal{U}(k)=\mathcal{U}_{\mathrm{nom}}(k)
	+\bar{\mathcal{U}}(k)+\widetilde{\mathcal{U}}(k),
\end{equation}
where $\mathcal{U}_{\mathrm{nom}}$ is the nominal current, $\bar{\mathcal{U}}$ is the balancing feedback designed in the following subsection, and $\widetilde{\mathcal{U}}$ is a feedforward compensation term. Define the combined disturbance term to be compensated as
\begin{equation}
	\mathcal{H}(k)=\mathcal{B}_{\Delta}\mathcal{U}_{\mathrm{nom}}(k)
	+\overline{\mathcal{A}}_{\Delta}\mathcal{Z}(k)
	+\mathcal{E}_{\Delta}\mathcal{D}(k).
\end{equation}
In the conventional feedforward formulation, exact cancellation is possible if $\mathcal{B}_{\Delta}$ is invertible or if $\operatorname{Range}\!\left(\begin{bmatrix}\overline{\mathcal{A}}_{\Delta}&\mathcal{E}_{\Delta}\end{bmatrix}\right)\subseteq\operatorname{Range}(\mathcal{B}_{\Delta})$. Under either condition, the standard pseudoinverse-based feedforward input $\widetilde{\mathcal{U}}_{\mathrm{ex}}(k)=-\mathcal{B}_{\Delta}^{\dagger}\mathcal{H}(k)$ satisfies $\mathcal{B}_{\Delta}\widetilde{\mathcal{U}}_{\mathrm{ex}}(k)+\mathcal{H}(k)=0$. In the present electro-thermal balancing problem, however, the pseudoinverse of $\mathcal{B}_{\Delta}$ is ill-conditioned because the ambient-temperature acts only through the surface-temperature dynamics. Hence, the range condition is not guaranteed, and the exact compensator may be infeasible or may produce excessively large currents when small singular values are present.

To address this limitation, the feedforward compensation is formulated as the Tikhonov-regularized problem
\begin{equation}
	\underset{\widetilde{\mathcal{U}}(k)}{\operatorname{min}}
	\left\{\frac{1}{2}\left\|\mathcal{B}_{\Delta}\widetilde{\mathcal{U}}(k)
	+\mathcal{H}(k)\right\|_2^2
	+\frac{\lambda}{2}\left\|\widetilde{\mathcal{U}}(k)\right\|_2^2\right\},
\label{regularized_feedforward_problem}
\end{equation}
whose unique solution is
\begin{equation}
	\widetilde{\mathcal{U}}(k)
	=-\left(\mathcal{B}_{\Delta}^{\top}\mathcal{B}_{\Delta}
	+\lambda I_{N_c N_u}\right)^{-1}
	\mathcal{B}_{\Delta}^{\top}\mathcal{H}(k),
\label{ff_controller}
\end{equation}
where $I_q$ denotes the $q\times q$ identity matrix and $\lambda>0$ is the regularization parameter. The regularization guarantees that the inverse in \eqref{ff_controller} exists and limits amplification along weakly actuated directions. Because regularization does not generally provide exact cancellation, define the residual
\begin{equation}
	\mathcal{R}_{\lambda}(k)
	=\mathcal{B}_{\Delta}\widetilde{\mathcal{U}}(k)+\mathcal{H}(k).
\label{regularization_residual}
\end{equation}
\begin{assumption}\label{assump:zero_regularization_residual}
The regularization parameter $\lambda$ is selected such that the compensation residual is negligible over the operating region considered in this study. Accordingly, $\mathcal{R}_{\lambda}(k)=0$ is assumed for controller synthesis and analysis.
\end{assumption}
Under Assumption~\ref{assump:zero_regularization_residual}, substitution into \eqref{global_error_dynamics} gives the compensated error dynamics
\begin{equation}
	\Delta(k+1)=\mathcal{A}_{\Delta}\Delta(k)
	+\mathcal{B}_{\Delta}\bar{\mathcal{U}}(k).
\label{compensated_error_dynamics}
\end{equation}
Thus, $\lambda$ provides a tradeoff between cancellation accuracy and feedforward-current magnitude. When $\mathcal{H}(k)\in\operatorname{Range}(\mathcal{B}_{\Delta})$, the residual approaches zero as $\lambda\rightarrow0^{+}$, which supports Assumption~\ref{assump:zero_regularization_residual} for a sufficiently small regularization parameter.

\subsection{Cell Balancing Controller Design}
As stated in \eqref{state_consensus}, the cell-balancing objective is to achieve consensus among neighboring cells in SoC and temperature. In this study, Koopman-based model predictive control (KMPC) is utilized to determine the balancing current $\bar{\mathcal{U}}(k)$ using the compensated error dynamics in \eqref{compensated_error_dynamics}. At each sampling instant, the KMPC problem is formulated as
\begin{subequations}\label{mpc_formulation}
\begin{align}
	\underset{\bar{\mathcal{U}}(\cdot|k)}{\operatorname{min}}\quad
	\sum_{j=0}^{N_p-1}&\left(
	\left\|\Delta(j|k)\right\|_{Q}^{2}
	+\left\|\bar{\mathcal{U}}(j|k)\right\|_{R}^{2}
	\right) + \left\|\Delta(N_p|k)\right\|_{P}^{2}, \label{mpc_cost}\\
	\text{s.t.}\quad
	&\Delta(j+1|k)=\mathcal{A}_{\Delta}\Delta(j|k)
	+\mathcal{B}_{\Delta}\bar{\mathcal{U}}(j|k), \label{mpc_prediction}\\
	&\Delta(0|k)=\Delta(k),\\
	&\bar{\mathcal{U}}_{\min}\leq\bar{\mathcal{U}}(j|k)
	\leq\bar{\mathcal{U}}_{\max},\quad j=0,\ldots,N_p-1,
\end{align}
\end{subequations}
where $N_p$ is the prediction horizon and $\|\bm{v}\|_{W}^{2}=\bm{v}^{\top}W\bm{v}$. The matrix $R\in\mathbb{R}^{N_cN_u\times N_cN_u}\succ0$ penalizes the balancing-current effort, whereas $Q,P\in\mathbb{R}^{(N_c-1)N_o\times(N_c-1)N_o}\succeq0$ are the stage and terminal weighting matrices, respectively. The vectors $\bar{\mathcal{U}}_{\min}$ and $\bar{\mathcal{U}}_{\max}$ denote the lower and upper bounds, respectively, on the balancing current. Only the first optimal control action is applied, and the optimization is repeated at the next sampling instant. Because \eqref{mpc_formulation} has a quadratic objective, linear prediction dynamics, and affine input constraints, it can be posed as a convex quadratic program. With $R\succ0$, every feasible instance has a unique global optimizer and can be handled efficiently using standard quadratic-programming solvers. In contrast, nonlinear MPC generally involves nonlinear prediction constraints and may also employ a nonlinear, potentially nonconvex objective, leading to a more computationally demanding online optimization problem \cite{ren2025koopman,allgower2012nonlinear}.

\begin{table}[b]
\caption{Parameters of the Cell-Specific Electro-Thermal Models}
\label{table_1}
\centering
\small
\setlength{\tabcolsep}{6pt}
\begin{tabular}{lccc}
\toprule
Parameter & Cell 1 & Cell 2 & Cell 3 \\
\midrule
$Q_c$ ($\mathrm{Ah}$) & 111.32 & 115.20 & 116.39 \\
$R_0$ ($\mathrm{m}\Omega$) & 0.48 & 0.64 & 0.33 \\
$R_1$ ($\mathrm{m}\Omega$) & 0.31 & 0.36 & 0.62 \\
$C_1$ ($\mathrm{kF}$) & 342.63 & 352.18 & 260.99 \\
$R_d$ ($\mathrm{K/W}$) & 1.71 & 1.77 & 1.67 \\
$C_d$ ($\mathrm{J}/{}^{\circ}\mathrm{C}$) & 65.06 & 63.06 & 60.52 \\
$R_v$ ($\mathrm{K/W}$) & 2.78 & 2.94 & 2.81 \\
$C_v$ ($\mathrm{J}/{}^{\circ}\mathrm{C}$) & 4.43 & 4.08 & 4.27 \\
\bottomrule
\end{tabular}
\end{table}

\section{Simulation Results and Comparison}
\subsection{Koopman Operator Model Validation}
The numerical study considers a pack of three series-connected cells ($N_c=3$). Before evaluating the closed-loop balancing performance, the prediction accuracy of the cell-specific Koopman models is examined to determine whether the nonlinear electro-thermal dynamics, including heat generation, are represented adequately in the lifted space. To construct the snapshot matrices in \eqref{snapshot}, the dynamics of each cell are integrated using a foward Euler method with a fixed sampling interval of $\Delta t=1~\mathrm{s}$. The cell-specific parameter values are provided in Table~\ref{table_1}.

\begin{figure}
    \centering
    \includegraphics[width=100mm]{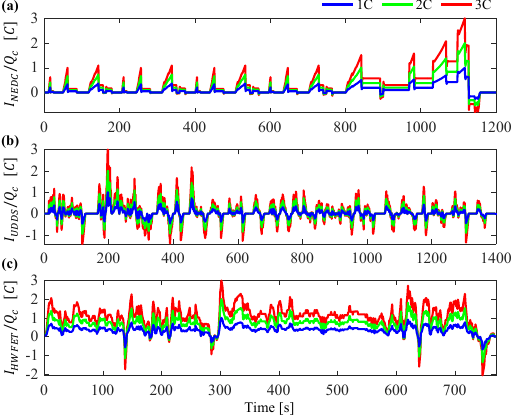}
    \caption{Current profiles used to train the Koopman models: (a) NEDC, (b) UDDS, and (c) HWFET.}
    \label{figure_2}
\end{figure}

The training inputs comprise the New European Driving Cycle (NEDC), Urban Dynamometer Driving Schedule (UDDS), and Highway Fuel Economy Test Cycle (HWFET), with each profile scaled to $1\mathrm{C}$, $2\mathrm{C}$, and $3\mathrm{C}$ (see Fig.~\ref{figure_2}). These combinations produce nine distinct excitation cases. For each cell and excitation case, 100 initial conditions are sampled at $t=0$, yielding 900 training rollouts per cell. For each rollout, the ambient temperature is held constant and set equal to the initial surface temperature.

A common lifting dictionary is used for all cells and consists of the original state coordinates augmented by three radial basis functions (RBFs), $\psi_{\ell}(\bm{x})=\exp\!\left(-\frac{\left\|\bm{x}-\bm{c}_{\ell}\right\|_2^2}{2\sigma^2}\right)$, where $\ell=1,2,3$ and $\bm{c}_{\ell}\in\mathbb{R}^{N_x}$ is the center of the $\ell$th RBF and $\sigma=0.5$ is the common width. The components of each RBF center $\bm{c}_{\ell}$ are sampled independently and uniformly from the interval $[-\pi/2,\pi/2]$. The lifting map is then $\bm{\Psi}(\bm{x})=\begin{bmatrix}\bm{x}^{\top} & \psi_1(\bm{x}) & \psi_2(\bm{x}) & \psi_3(\bm{x})\end{bmatrix}^{\top}$, where $\bm{\Psi}(\bm{x})\in\mathbb{R}^{N_o}$ and $N_o=N_x+3=7$. Because the physical states occupy the first $N_x$ entries of the lifted state, the reconstruction matrix is selected as $C_i=C=\begin{bmatrix}I_{N_x} & 0_{N_x\times3}\end{bmatrix}\in\mathbb{R}^{N_x\times N_o}$.

\begin{figure}[t]
    \centering
    \includegraphics[width=100mm]{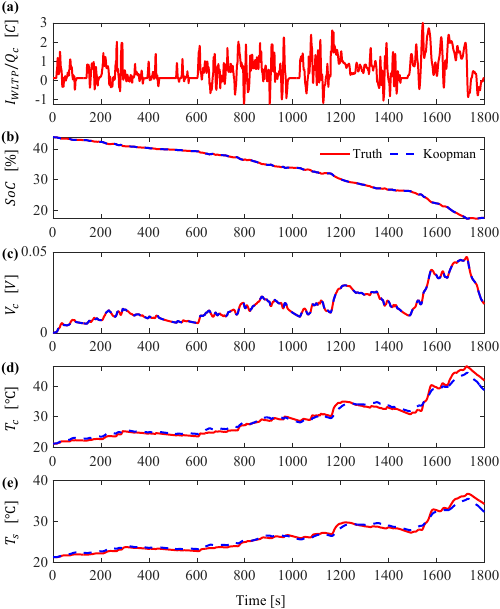}
    \caption{Validation of the Koopman model against the nonlinear electro-thermal simulation under the WLTP current profile: (a) WLTP input current, (b) $SoC$, (c) $V_c$, (d) $T_c$, and (e) $T_s$.}
    \label{figure_3}
\end{figure}

Figure~\ref{figure_3} compares the Koopman-model predictions with the nonlinear electro-thermal simulation under a Worldwide Harmonized Light Vehicles Test Procedure (WLTP) current profile that was excluded from the training dataset. The predicted SoC and $V_c$ trajectories closely match the corresponding simulated responses, consistent with the linear structure of the electrical subsystem. The temperature predictions exhibit modest discrepancies because the finite-dimensional Koopman model approximates the nonlinear heat-generation dynamics. The root-mean-square errors (RMSEs) for the core and surface temperatures are $1.01~{}^{\circ}\mathrm{C}$ and $0.61~{}^{\circ}\mathrm{C}$, respectively. These validation results indicate that the identified Koopman model provides sufficient prediction accuracy over the tested trajectory for use in the balancing-controller design.

\begin{figure}
    \centering
    \includegraphics[width=100mm]{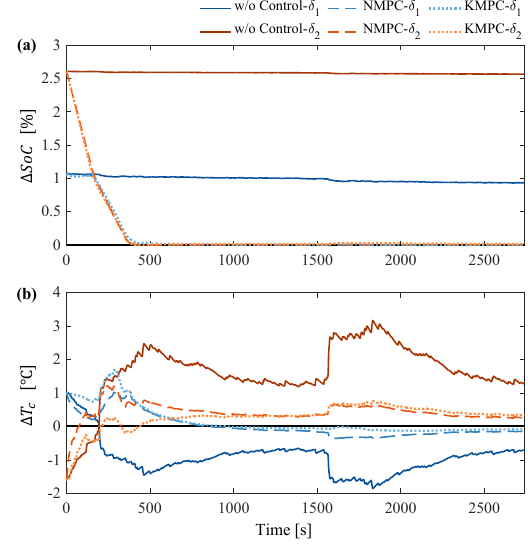}
    \caption{Cell-balancing performance over two consecutive UDDS cycles: (a) neighboring-cell SoC differences and (b) neighboring-cell core-temperature differences.}
    \label{figure_4}
\end{figure}

\subsection{Simulation setup}
Nonlinear model predictive control (NMPC) is employed as a benchmark for assessing the proposed KMPC controller. Its prediction model is constructed directly from the nonlinear electro-thermal dynamics in \eqref{electrothermalmodel} and \eqref{thermalmodel}, with the neighboring-cell physical-state errors stacked analogously to the lifted errors in \eqref{global_error_dynamics}. Unlike the Koopman model, the NMPC model retains the nonlinear heat-generation term $Q_{gen,i}$ and the resulting state--input coupling. For a consistent comparison, both controllers use the same finite-horizon cost structure, terminal penalty, and balancing-current constraints. However, the resulting optimization problem in NMPC is nonlinear.

For both KMPC and NMPC, the prediction horizon is set to $N_p=10$, and the componentwise balancing-current limits are $-20~\mathrm{A}$ and $20~\mathrm{A}$. Define the physical-state weighting block as $Q_x=P_x=\operatorname{diag}(10000,100,100,100)$. The NMPC stage and terminal weighting matrices are $Q=P=I_{N_c-1}\otimes Q_x$. For KMPC, the corresponding lifted-state weighting block is $Q_z=P_z=\operatorname{diag}(10000,100,100,100,0,0,0)$ and the global weighting matrices are $Q=P=I_{N_c-1}\otimes Q_z$. Hence, both controllers apply identical penalties to the physical cell-to-cell errors, while the three additional RBF observables in the KMPC model are unpenalized. The balancing-current weighting matrix is selected as $R=0.1 \times I_{N_cN_u}$ for both controllers.

Both optimization problems are implemented in YALMIP \cite{lofberg2004yalmip}. The nonlinear NMPC problem is solved using MATLAB's \texttt{fmincon}, whereas the convex KMPC quadratic program is solved using \texttt{quadprog}. All simulations are performed on a desktop computer equipped with a 3.0-GHz Apple M4 processor and 16~GB of RAM.

\subsection{Balancing results}
The balancing controllers are evaluated over two consecutive UDDS cycles. The initial SoCs are set to $[SoC_1(0),SoC_2(0),SoC_3(0)]^{\top}=[86.22,85.15,82.54]^{\top}\%$. Each cell is assumed to be internally isothermal initially, while the initial temperatures differ among cells; specifically, $[T_{c,1}(0),T_{c,2}(0),T_{c,3}(0)]^{\top}=[T_{s,1}(0),T_{s,2}(0),T_{s,3}(0)]^{\top}=[27.66,26.64,28.20]^{\top}~{}^{\circ}\mathrm{C}$.

Figure~\ref{figure_4} presents the neighboring-cell SoC and $T_c$ differences, defined as $\Delta SoC_i=SoC_i-SoC_{i+1}$ and $\Delta T_{c,i}=T_{c,i}-T_{c,i+1}$ for $i=1,2$. The solid, dashed, and dotted curves in Fig.~\ref{figure_4} correspond to the uncontrolled, NMPC, and KMPC cases, respectively. Without balancing control, the SoC differences remain near $1.0\%$ and $2.6\%$ for the first ($\delta_1$) and second cell ($\delta_2$) pairs, respectively. Such imbalance can cause one cell to reach its charge or discharge limit before the remaining cells, thereby reducing the usable pack capacity and promoting nonuniform degradation. Relative to the zero error target, the $\Delta SoC$ RMSEs for the first and second cell pairs are $0.30\%$ and $0.48\%$ with NMPC and $0.34\%$ and $0.50\%$ with KMPC. Relative to the uncontrolled case, KMPC reduces the RMSEs of the neighboring-cell SoC differences by $66$--$81\%$.

For the uncontrolled case, the $\Delta T_{c}$ RMSEs are $1.07~{}^{\circ}\mathrm{C}$ and $1.85~{}^{\circ}\mathrm{C}$ for the first and second cell pairs, respectively. Both controllers substantially reduce these temperature differences. The corresponding RMSEs are $0.34~{}^{\circ}\mathrm{C}$ and $0.50~{}^{\circ}\mathrm{C}$ with NMPC and $0.46~{}^{\circ}\mathrm{C}$ and $0.45~{}^{\circ}\mathrm{C}$ with KMPC. Relative to the uncontrolled case, KMPC reduces the RMSEs of the neighboring-cell SoC differences by $57$--$76\%$. Overall, KMPC achieves balancing performance comparable to that of NMPC while requiring less online computation. The computation times per optimization are $0.0028~\mathrm{s}$ for KMPC and $0.0259~\mathrm{s}$ for NMPC; thus, the NMPC solve time is approximately $9.25$ times that of KMPC.

\section{Conclusion}
This study presented a Koopman-operator-based control framework for simultaneous SoC and temperature balancing of heterogeneous series-connected lithium-ion cells. Cell-specific Koopman predictors were identified from nonlinear electro-thermal data and combined to construct global neighboring-cell error dynamics. The Tikhonov-regularized feedforward compensator was introduced to avoid direct use of the potentially ill-conditioned pseudoinverse and provided a numerically well-posed compensation law for the nominal current, parameter mismatch, and thermal disturbance terms. A constrained KMPC controller subsequently regulated the remaining errors. Simulation results demonstrated that the identified predictors reproduced the electrical and thermal responses under an unseen current profile with sufficient accuracy for control. Under two consecutive UDDS cycles, KMPC achieved SoC and core temperature balancing performance comparable to NMPC. However, the linear lifted prediction model reduced the computation time per optimization from $0.0259~\mathrm{s}$ for NMPC to $0.0028~\mathrm{s}$ for KMPC.

Future work will pursue four directions. First, the present study's assumption was that all electro-thermal states are directly available. In practical battery management systems, states such as SoC and core temperature are not directly measurable and must be reconstructed using an estimator. A robust controller design is required to explicitly accounts for state estimation errors as well as parameter uncertainty and modeling errors arising from the Koopman model. Second, the controller will be integrated with a battery degradation model to quantify and regulate the long term effects of balancing actions on cell aging. Third, the framework will be evaluated under a broader range of ambient temperatures, cycling profiles, and high and low SoC operating conditions. Fourth, the balancing controller will be coordinated with the pack heating, ventilation, and air conditioning system to jointly manage electrical imbalance and pack-level thermal conditions.

\nocite{*}
\bibliographystyle{asmejour}
\bibliography{references}

\end{document}